# Zero-shot Dynamic MRI Reconstruction with Global-to-local Diffusion Model

Yu Guan, Kunlong Zhang, Qi Qi, Dong Wang, Ziwen Ke, Shaoyu Wang, Dong Liang,
*Senior Member, IEEE*, Qiegen Liu, *Senior Member, IEEE*

*Abstract*—Diffusion models have recently demonstrated considerable advancement in the generation and reconstruction of magnetic resonance imaging (MRI) data. These models exhibit great potential in handling unsampled data and reducing noise, highlighting their promise as generative models. However, their application in dynamic MRI remains relatively underexplored. This is primarily due to the substantial amount of fully-sampled data typically required for training, which is difficult to obtain in dynamic MRI due to its spatio-temporal complexity and high acquisition costs. To address this challenge, we propose a dynamic MRI reconstruction method based on a time-interleaved acquisition scheme, termed the Global-to-local Diffusion Model. Specifically, fully encoded full-resolution reference data are constructed by merging under-sampled k-space data from adjacent time frames, generating two distinct bulk training datasets for global and local models. The global-to-local diffusion framework alternately optimizes global information and local image details, enabling zero-shot reconstruction. Extensive experiments demonstrate that the proposed method performs well in terms of noise reduction and detail preservation, achieving reconstruction quality comparable to that of supervised approaches.

## I. INTRODUCTION

Dynamic magnetic resonance imaging (MRI) is frequently employed to monitor dynamic anatomical processes, such as cardiac motion, by capturing a sequence of images at a high frame rate [1]. Nevertheless, the inherent physics of the image acquisition process, along with physiological limitations, impose constraints on the speed of MRI acquisition. Prolonged scan durations further complicate the imaging of moving structures [2]. One of the primary challenges in dynamic MRI is the reconstruction of high-dimensional images from sparse k-space data, which is sampled below the Nyquist rate. Consequently, it is imperative to accelerate dynamic MRI acquisition while maintaining superior image quality.

This work was supported in part by the National Key Research and Development Program of China under Grant 2023YFF1204300 and Grant 2023YFF1204302, in part by the National Natural Science Foundation of China under Grant 62122033. Y. Guan and K. Zhang are co-first authors. (Corresponding authors: Dong Liang, Qiegen Liu.)

Y. Guan is with School of Mathematics and Computer Sciences, Nanchang University, Nanchang 330031, China. (guanyu@email.ncu.edu.cn).

K. Zhang, Q. Qi, D. Wang, S. Wang and Q. Liu are with the Department of Electronic Information Engineering, Nanchang University, Nanchang 330031, China. ({zhangkunlong, qiqi, wangdong}@email.ncu.edu.cn, shaoyuwang22@gmail.com, liuqiegen@ncu.edu.cn).

Ziwen Ke is with Beckman Institute for Advanced Science and Technology, University of Illinois at Urbana-Champaign, Urbana, Illinois, USA.

D. Liang are with Paul C. Lauterbur Research Center for Biomedical Imaging, SIAT, Chinese Academy of Sciences, Shenzhen 518055, China (sophiasswang@hotmail.com, dong.liang@siat.ac.cn).

Numerous methods have been proposed for accelerated imaging with sparse sampling, including parallel imaging [3] and compressed sensing (CS) [4], [5]. These techniques expedite scan times by sampling only a limited number of phase encodings and then leveraging prior information to restore the missing encodings during the reconstruction phase. For instance, the k-t SPARSE algorithm [6], initially proposed by Lustig *et al.*, was designed for dynamic cardiac cine MRI. The method capitalizes on the sparsity in the time Fourier transform and spatial wavelet transform domains. Among CS-based techniques, k-t FOCUSS [7] exploits prediction and residual encoding, significantly sparsifying and estimating residual signals from the under-sampled k-t samples, thereby achieving high-quality image reconstruction. Combinations of CS with low-rank (LR) matrix completion schemes and spatio-temporal partial separability [8]-[10] have also been proposed to exploit correlations between the temporal profiles of the voxels, such as k-t SLR [8]. Recent approaches [11], [12] have utilized patch-based regularization frameworks to exploit geometric similarities in the spatio-temporal domain. These methods have made significant advancements in dynamic imaging, achieving improved results. However, they typically rely on manually designed priors and require tuning of hyperparameters to achieve optimal performance, which may not generalize well to test settings.

More recently, deep learning (DL) has revolutionized the image reconstruction field, rapidly becoming the state of the art. Inspired by the development of DL techniques across various imaging modalities [13]-[15], supervised learning approaches have been applied to the fast and accurate reconstruction of partially sampled MRI [16]-[21]. Several early approaches [22], [23] proposed using a single feed-forward convolutional neural network (CNN), such as SRCNN [24] and U-Net [25], to learn an end-to-end mapping between the observed k-space and the fully-sampled data in the image domain, thereby achieving better reconstruction results. However, these methods heavily rely on the quality and characterization of the training dataset. If the training data does not adequately cover the distribution of test data, the model's generalization performance will be limited. Model-based DL methods have developed rapidly, and several representative networks have emerged. For example, ADMM-Net [26] is defined on a data flow graph derived from the iterative process of the alternating direction method of multipliers algorithm and is used to optimize the MRI model based on CS. Similarly, VN-Net [27] has also attracted much attention. This network combines the mathematical structure of the variational model with DL and learns all parameters through offline training to achieve high-quality image reconstruction. In addition, Learned PD [28] iteratively extends the primal dual hybrid gradient algorithm to a learnable deep network architecture and gradually relaxes the constraints to reconstruct MR images from highly under-sampled k-space data. While these methods have achieved promising results in MRI, they require large amounts of fully-sampled data to improve reconstruction quality. In practical imaging tasks, the available training data is often limited.

In comparison, diffusion models have gained considerable interest as a novel class of generative model, as they are capable of providing a more precise representation of the data distribution and a remarkably high level of sample quality [29]-[33]. Jalal *et al.* [34] have conducted groundbreaking research by training diffusion models on magnetic resonance images, effectively employing these models as prior information in the inversion process. This approach improves the reconstruction capability of realistic MRI data and provides a new reference in the field of medical imaging. Subsequently, a similar approach was applied to accelerated MRI by Chung *et al.* [35], who trained a continuous time-dependent score function using denoising score matching to achieve high accuracy reconstruction. Recently, Yu *et al.* [36] proposed a k-space and image dual-domain collaborative universal generative model, which combines the score-based prior with a LR regularization penalty to reconstruct highly under-sampled measurements. Nevertheless, all of the above methods for dynamic MRI reconstruction based on diffusion models use fully-sampled data to train the models, and these researches still have capacity to improve in recovering fine details or structures.

Given the difficulty of acquiring fully-sampled data in dynamic MRI reconstruction, researchers have also proposed various methods to address this issue. One notable approach is the global interpolation network [37], which learns global dependencies among low- and high-frequency components of 2D+t k-space and uses them to interpolate unsampled data. Nevertheless, this approach depends on local operators such as convolution, which constrains the flexibility of the scanning process. Another significant approach is the time-interleaved acquisition scheme [38], which eliminates the need for separately acquiring additional reference data. Signals from directly adjacent time frames can be merged to build a set of fully encoded full-resolution reference data for coil calibration, enabling full image acceleration. Specifically, with each acquired time frame in a series, a new set of autocalibration signal (ACS) data can be used to reconstruct the next frame in the time series, thereby tracking changes in relative coil sensitivities over time, greatly enhancing flexibility when scanning dynamic MRI data. Although this time-interleaved acquisition scheme has achieved certain results, it reduces the temporal resolution.

Motivated by the abovementioned works, we designed a novel dynamic MRI reconstruction principle that inherits the advantages of previous methods without compromising temporal resolution. The time-interleaved acquisition scheme is integrated with the Global-to-local Diffusion Model to achieve zero-shot dynamic MRI reconstruction, referred to as GLDM. Specifically, the proposed framework constructs two training datasets using a time-interleaved acquisition scheme: one obtained by directly merging all frames to fully leverage each frame, and the other composed of local frame merging to retain more detailed information of each frame. Additionally, traditional LR constraints are employed as a secondary priori regularization to leverage sparsity in both the spatial and temporal domains. Overall, GLDM combines score-based generative models with traditional methods into a new constrained optimization term in the reconstruction stage, which is then solved through an alternating minimization scheme. The main contributions and observations of this work are summarized as follows:

- ***Zero-shot learning via a time-interleaved acquisition scheme.*** Two different time-interleaved acquisition schemes are exploited to generate fully encoded data for network training, eliminating the need for fully-sampled data while maintaining reconstruction accuracy, thereby achieving zero-shot learning. In addition, the influence of different time frame merging strategies on model performance is systematically analyzed to identify the optimal configuration. This optimization enhances both the efficiency of model training and the quality of the resulting reconstructions.
- ***Two-stage iterative with global-to-local diffusion model.*** The global-to-local diffusion model framework is utilized to optimize the global structures and local details of the image alternatively. Additionally, the integration of LR operators and data consistency (DC) modules strengthens the robustness of the model, facilitating optimal reconstruction performance. By synthesizing these components within a unified model framework, the proposed method achieves reconstruction quality comparable to supervised learning.

The remainder of this paper is organized as follows. In Section II, we briefly introduce the forward model of dynamic MRI and review the background knowledge of k-space data and cascading models. Section III explains our proposed method GLDM in detail, which describes the global-to-local forward diffusion process, including two different time-interleaved acquisition schemes to construct the training dataset and a two-stage learning mechanism approach. The performance of the proposed reconstruction method is demonstrated and compared with classic and DL methods in Section IV. Finally, discussions and conclusions are presented in Sections V and VI, respectively.

## II. PRELIMINARIES

### A. Forward Imaging Model

The forward model of dynamic MRI can be expressed by the following formula:

$$y = Ak + \epsilon, \quad (1)$$

where $k \in \mathbb{C}^{N_x \times N_y \times N_t}$ represents the k-space data to be reconstructed, $y \in \mathbb{C}^{N_x \times N_y \times N_t}$ is the multi-coil under-sampled k-space data, where $N_x$ and $N_y$ are the image height and width, and $N_t$ is the number of time frames. $A$ is the measurement matrix determined according to $k$, $\epsilon$ represents the noise of the measurement. More specifically, $A = F_u S$ is the multi-coil dynamic MRI system matrix, $S$ is the coil sensitivity, and $F_u$ is the under-sampled Fourier transform. Due to insufficient information of $k$, the problem of recovering $k$ from $y$ is severely ill-posed, which precludes direct inversion of Eq. (1). To address this, regularization techniques are employed, resulting in the following optimization problem:

$$\min_{k}\{\| Ak - y \|_2^2 + \lambda \mathcal{R}(k)\}, \quad (2)$$

where $\| Ak - y \|_2^2$ is the data fidelity term, which ensures that the reconstructed k-space data $k$ is consistent with the actual measurement value $y$. $\mathcal{R}(k)$ represents a regularization function that forces a priori on $k$, with $\lambda$ controlling the balance between fidelity and regularization.

### B. K-space Interpolation Manner and Time-interleaved Acquisition Scheme

In dynamic MRI, due to limited scanning time, only a limited amount of k-space data can be obtained in each time frame. To address this challenge, researchers have explored various methods for processing k-space data, including k-space interpolation and time-interleaved acquisition schemes. The k-space interpolation method can avoid the fundamental mismatch between the real continuous image and the discrete grid singular points [39]. Inserting missing k-space data primarily relies on various physical priors of MR images. The most classic example [40], through statistical observation or transformation of image domain sparsity in the Fourier domain, found that k-space data can be linearly predicted in the neighborhood. Cui *et al.* [41] proposed a guaranteed k-space interpolation method for MRI, combining a neural network to fully utilize the three physical priors of MRI to achieve satisfactory reconstruction results. Du *et al.* [42] proposed an adaptive CNN algorithm for k-space data interpolation. The algorithm adopts a residual encoder-decoder network architecture to interpolate under-sampled k-space data by taking spatially continuous slices as multi-channel inputs and k-space data from multiple coils.

In parallel, the time-interleaved acquisition scheme has emerged as a significant area of research, closely related to the methodology presented in this paper. A detailed illustration of the time-interleaved acquisition scheme is provided in Fig. 1. Breuer *et al.* [38] proposed to combine the time-interleaved acquisition scheme with GRAPPA, directly merging the signals of adjacent time frames to construct a set of fully encoded full-resolution reference data for coil calibration, thereby improving acquisition efficiency. Ke *et al.* [43] further developed an unsupervised DL method for multi-coil cine MRI through a time-interleaved acquisition scheme. The fully encoded full-resolution reference data constructed by this method is used to train a parallel network to reconstruct the image of each coil. The success of the above work proves that the time-interleaved acquisition scheme can effectively accelerate imaging and guarantee image quality to a certain extent in the field of dynamic MRI.

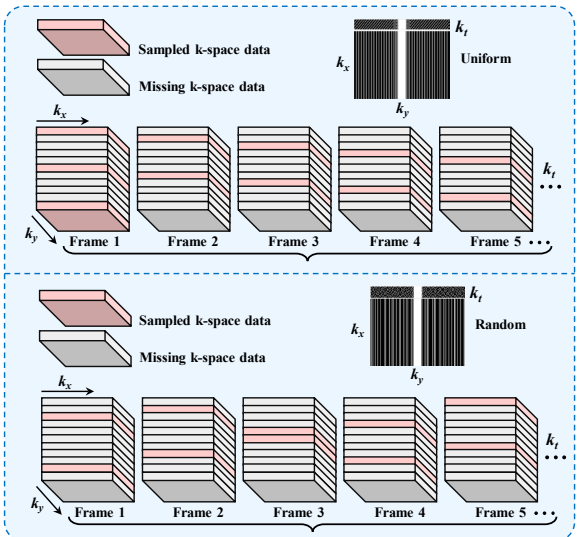

**Fig. 1.** Time-interleaved acquisition scheme. The core of the approach is to construct a complete k-space dataset by merging any number of adjacent time frames. In the above example, two different under-sampled patterns (uniform and random) at 5-fold acceleration are acquired via a time-interleaved acquisition scheme.

*C. Multi-model Strategy for Cascading Refinement*

In MRI reconstruction, models trained on a specific dataset are able to effectively capture the unique features of the data. Notably, the cascaded model shows significant advantages over a single model by focusing on different data features. Existing work has revealed that multi-model refinement models are effective in various medical image processing tasks. For instance, Tu *et al.* [44] combined information from both the k-space and image domains, enhancing the quality and accuracy of parallel MRI reconstruction through the synergy of two models and multiple iterations. Similarly, Zhou *et al.* [45] introduced a dual-domain self-supervised transformer approach, improving the precision and efficiency of multi-contrast MRI reconstruction by applying self-supervised learning in both the image and k-space domains. In the domain of medical image segmentation, iterative refinement model frameworks also proved to achieve higher accuracy and efficiency. In the field of medical image segmentation, Kutra *et al.* [46] proposed a framework that fully leveraged the strengths of different models by integrating their segmentation outputs through a fusion strategy, enhancing segmentation accuracy and reliability. Furthermore, Lyu *et al.* [47] presented a two-stage cascade model for MRI brain tumor segmentation, where the model first learned general image features using variational autoencoders, followed by attention gates that dynamically assigned weights to different regions based on the learned features, more effectively capturing tumor characteristics in MRI.

The aforementioned multi-model refinement methods have achieved remarkable success, substantiating the effectiveness and superiority of iterative refinement strategies. Thus, it can be inferred that multi-model cascade refinement is particularly well suited to address intricate challenges in MRI reconstruction.

### III. PROPOSED METHOD

*A. Motivation*

Dynamic MRI data acquisition is usually limited by long acquisition cycles and motion artifacts. In traditional methods, due to limited scanning time, only limited k-space data can be collected in each time frame, resulting in reduced temporal resolution and aggravated motion artifacts. Nevertheless, the sampled k-space data can still provide effective and reliable information, which is of great value for clinical diagnosis. Therefore, how to balance acquisition efficiency and image reconstruction quality in dynamic MRI has become a pressing challenge.

To address this issue, we introduce a time-interleaved acquisition scheme to enhance temporal resolution and mitigate motion artifacts by offsetting k-space data in the time domain. In addition, in order to strike a balance between global structural features and fine local details, the proposed framework adopts global and local time-interleaved acquisition schemes. The global scheme provides a broader structural perspective by sampling less data at more time points to ensure that the overall image structure is fully captured; in contrast, the local scheme collects more data at fewer time points and focuses on preserving local details. The combined scheme can generate high-quality full-resolution reference data.

To effectively reconstruct high-quality images from these acquisition schemes, we leverage the powerful probabilistic modeling capabilities of diffusion models. Diffusion models significantly improve the quality of reconstructed images by learning the compositional structure or specific prior distribution of samples. Specifically, we propose a two-stage iterative refinement diffusion strategy to adapt to the characteristics of global and local time-interleaved schemes. Utilizing score-based diffusion models, we train them to learn the prior distribution of image features. During the iterative reconstruction process, the model is conditionally sampled by alternating the execution of the stochastic differential equations (SDE) solver, the LR constraint, and the DC step, ensuring an optimal reconstruction output. Fig. 2 illustrates the overall workflow of the proposed method. The following sections will detail the prior learning process and iterative reconstruction scheme of GLDM.

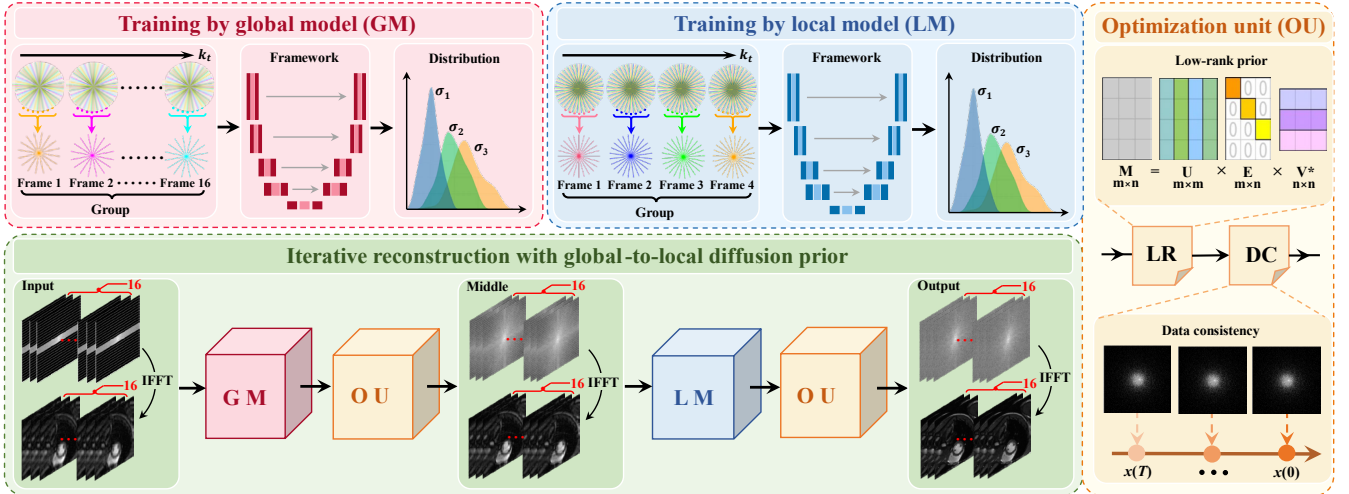

**Fig. 2.** The schematic of the proposed GLDM algorithm. Red and blue parts represent the training stage that fully encoded full-resolution reference data is constructed through a time-interleaved acquisition scheme. Red part merges all time frames to train the global model (GM) while the blue part merges local time frames to train the local model (LM). Green part represents the reconstruction stage which the structure of the reconstruction model exists in a cascade form and the under-sampled k-space data (16 frames) are sequentially input into the network. At the same time, optimization unit (OU) containing a LR operator and a DC term is introduced to better remove aliasing and restore details.

*B. Global-to-local Forward Diffusion Process*

***Global and local time-interleaved acquisition schemes***: We have retrospectively under-sampled each original fully-sampled k-space dataset according to a uniform time-interleaved sampling pattern. This approach is distinctive in that it has the capacity to markedly improve the utilization of information from under-sampled frames while maintaining contrast information that is highly similar to that of fully-sampled frames. Fig. 3 illustrates a schematic diagram of the local time-interleaved acquisition scheme with an acceleration factor of $R = 4$. Specifically, for each frame, the frequency encoding (along $k_x$) is fully sampled, while the phase encoding (along $k_y$) is uniformly under-sampled. It is noteworthy that the central phase encoding of the four frames remains fully sampled. By combining the four under-sampled frames into a temporary frame, whose central phase encoding encompasses data from all under-sampled frames while the remaining phase encoding contains data from a single under-sampled frame, a full representation is achieved. Subsequently, the central phase encodings of the four under-sampled frames are utilized to ensure data fidelity of the temporary frame, ensuring its contrast information closely resembles that of a fully-sampled frame. Consequently, a fully encoded k-space training dataset ($k_l$) comprising local information is obtained. Similarly, by setting the acceleration factor of $R = 16$, the global time-interleaved acquisition scheme obtains a fully encoded k-space training dataset ($k_g$) containing global information. In this global scheme, the overall structural information of the image can be captured, avoiding the information loss caused by the local scheme. The proposed time-interleaved acquisition scheme allows us to obtain the same number of fully encoded datasets with richer features by processing only under-sampled data, without increasing the actual acquisition time, which is of great importance for training models for DL.

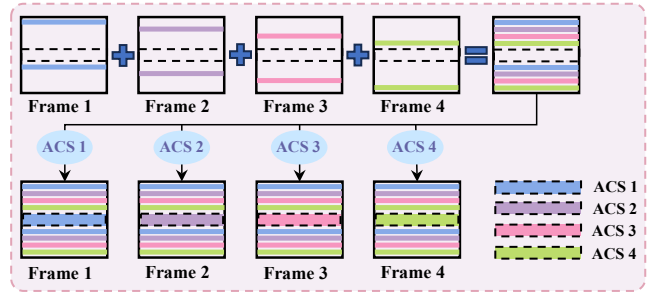

**Fig. 3.** The time-interleaved acquisition scheme of 4 frames of dynamic MRI is shown. The ACS of each frame remains unaltered, while the remainder of the area is filled with data from adjacent frames. The distinct colors represent data contributions from different frames.

***Two-stage learning mechanism***: The two types of training datasets obtained above are fed into a two-stage training network. Although these training datasets are synthetic, they effectively represent real fully-sampled data. In the first stage, we train a global model (GM) using the global time-interleaved acquisition dataset. Although this method effectively enhances the signal-to-noise ratio and captures the overall distribution characteristics of the data, it inevitably sacrifices certain important details. To solve this problem, In the second stage, we train a local model (LM) using the local time-interleaved acquisition dataset, focusing on extracting local features and operates alternately with GM in an iterative optimization process. In this way, the model achieves improved accuracy and a refined ability to reconstruct intricate details, creating a complementary relationship between global and local learning. The prior learning process in the score-based model is accomplished by using SDE to transform the complex data distribution into a known prior distribution. The specific prior learning process is as follows:

***Learning global prior information***: The prior learning process of the score-based diffusion model operates by using

SDE to transform the complex data distribution into a predefined prior distribution. The objective is to establish a diffusion process $\{k_g(t)\}_{t=0}^T$ indexed by a continuous time variable $t \in [0, T]$, where $k_g(0) \sim p_0$ represents the initial distribution of a sample dataset, and $k_g(T) \sim p_T$ represents a tractable prior distribution for efficient sample generation. The general form of this process can be described as follows:

$$dk_g = f(k_g, t)dt + g(t)dw, \quad (3)$$

where $w$ is the standard Wiener process (also known as Brownian motion), $f(k_g, t)$ is the drift coefficient function, which describes the deterministic changes of the system, and $g(t)$ is the diffusion coefficient function, which describes the random changes of the system. It is worth noting that the SDE has a strong unique solution. In general, different $f(\cdot)$ and $g(\cdot)$ functions can be selected to design the SDE in Eq. (3) to diffuse the data distribution into a fixed prior distribution.

Following the work of Song et al. [48], we adopt the form of variance exploding SDE (VE-SDE), which usually leads to higher sample quality. Here we set $f(k_g, t) = 0$ and choose $g(t) = \sqrt{d[\sigma^2(t)]/dt}$. Thence, the forward VE-SDE can be expressed as:

$$dk_g = \sqrt{\frac{d[\sigma^2(t)]}{dt}} dw, \quad (4)$$

where $\sigma(t)$ is regarded as a Gaussian noise function in continuous time $t \in [0, T]$. By introducing increasing noise in the k-space domain, VE-SDE gradually transforms the data distribution into a known prior distribution. The inverse VE-SDE restores the prior distribution to the original data distribution by denoising. The inverse VE-SDE can be expressed as follows:

$$dk_g = -\left(\sqrt{\frac{d[\sigma^2(t)]}{dt}}\right)^2 \nabla_{k_g} \log p_t(k_g)dt + \sqrt{\frac{d[\sigma^2(t)]}{dt}} d\bar{w}, \quad (5)$$

where $\bar{w}$ is the standard Wiener process when time flows backward from $T$ to $0$, and $dt$ is an infinitesimal negative time step. For the gradient $\nabla_{k_g} \log p_t(k_g)$ of the marginal distribution at each time $t$, we can derive the reverse diffusion process from Eq. (5) and simulate it into the distribution sampling of $p_0$. By training the scoring network based on VE-SDE, the gradient of the k-space data distribution can be estimated. The optimization objective of the model is formulated as:

$$\theta_{k_g}^* = arg \min_{\theta_{k_g}} \mathbb{E}_t\{\lambda_{k_g}(t)\mathbb{E}_{k_g(0)}\mathbb{E}_{k_g(t)|k_g(0)} \left[\left\|S_{\theta_{k_g}}(k_g(t), t) - \nabla_{k_g(t)} \log p_{0t}(k_g(t)|k_g(0))\right\|_2^2\right]\}, \quad (6)$$

where $\lambda_{k_g}(t)$ is a positive weight coefficient at time $t$, which is used to adjust the loss contribution at different time steps. $S_{\theta_{k_g}}(k_g(t), t)$ is the estimated score function used to predict the Gaussian noise at time $t$ and state $k_g(t)$. $\nabla_{k_g(t)} \log p_{0t}(k_g(t)|k_g(0))$ is the gradient of the exact score function. The optimization objective $\theta_{k_g}^*$ identifies the best neural network parameters $\theta_{k_g}$ by minimizing the loss function. The score network $S_{\theta_{k_g}}(k_g(t), t)$ is trained to accurately approximate the gradient of the data distribution, which enables the model to effectively capture and utilize global prior information by denoising and generating new data samples $p_0^*$ from the prior distribution $p_T$ during reconstruction.

***Learning local prior information***: In this stage, we use a similar learning method as in the global stage, but the focus shifts to capturing local details in the data. This stage aims to improve the quality and details of the generated samples in local areas by refining the model. The diffusion process is still modeled in the k-space domain. The following is the solution of VE-SDE:

$$dk_l = \sqrt{\frac{d[\sigma^2(t)]}{dt}} dw. \quad (7)$$

Similarly, we train another score-based model specifically for local data by solving the following optimization problem:

$$\theta_{k_l}^* = arg \min_{\theta_{k_l}} \mathbb{E}_t\{\lambda_{k_l}(t)\mathbb{E}_{k_l(0)}\mathbb{E}_{k_l(t)|k_l(0)} \left[\left\|S_{\theta_{k_l}}(k_l(t), t) - \nabla_{k_l(t)} \log p_{0t}(k_l(t)|k_l(0))\right\|_2^2\right]\}, \quad (8)$$

where $S_{\theta_{k_l}}(k_l(t), t)$ is the trained denoising score probability model, which is used to predict the noise at time $t$ and state $k_l(t)$. This optimization objective ensures that the scoring network can accurately estimate the local data distribution.

### C. Optimization Refinement Reconstruction

In the previous section, the process of prior information learning is systematically explained, laying the foundation for the generation and optimization of global and local models. Through this learning process, two complementary models are successfully trained to work together in the reconstruction process. In this section, we delve into the details of the key steps in the image reconstruction process to further demonstrate its effectiveness in MRI reconstruction.

During the sampling process, we alternate between predictors and correctors. The inverse diffusion predictor is used for data generation, and the corrector further optimizes the generated data. The predictor can be any reverse time SDE numerical solver with a fixed discretization strategy. The corrector adopts a modified version of annealed Langevin dynamics to improve interpretability and empirical performance. The method is based on the reverse process of the diffusion model. Specifically, given the initial data $k_{t+1}$ (the state at time step $t + 1$), GM denoises and generates $k_{g,t}$ via the following update formula:

$$\begin{cases} k_{g,t} = k_{g,t+1} + (\sigma_{t+1}^2 - \sigma_t^2)S_{\theta_{k_g}}(k_{g,t+1}, t+1) + \sqrt{\sigma_{t+1}^2 - \sigma_t^2}z_{g,t} \\ k_{g,t} = k_{g,t} + \epsilon_t s_{\theta_{k_g}}(k_{g,t}, t) + \sqrt{2\epsilon_t}z_{g,t} \end{cases}, \quad (9)$$

where the first line of formulae is the predicting step and the second line is the correcting step. $S_{\theta_{k_g}}(k_{g,t+1}, t+1)$ represents GM function, which is used to predict or estimate the gradient of the data distribution at time $t + 1$, $(\sigma_{t+1}^2 - \sigma_t^2)$ is the change in noise variance, which adjusts the amplitude of the model output, and $\sqrt{\sigma_{t+1}^2 - \sigma_t^2}z_{g,t}$ represents the injected noise, which simulates the randomness in the diffusion process. $\epsilon_t s_{\theta_{k_g}}(k_{g,t}, t)$ represents the prediction correction of the model at the current time step. $\sqrt{2\epsilon_t}z_{g,t}$ represents the noise added during the correction process to ensure the randomness and diversity of the generation process. After generating the global data, we further refine the generated samples using the LM, which ensures the generated data $k_{l,t}$ more accurately conform to the target distribution. The update process for the LM follows a similar structure, using the following formula

$$\begin{cases} k_{l,t} = k_{l,t+1} + (\sigma_{t+1}^2 - \sigma_t^2)S_{\theta_{k_l}}(k_{l,t+1}, t+1) + \sqrt{\sigma_{t+1}^2 - \sigma_t^2}z_{l,t} \\ k_{l,t} = k_{l,t} + \epsilon_t S_{\theta_{k_l}}(k_{l,t}, t) + \sqrt{2\epsilon_t}z_{l,t} \end{cases}, \quad (10)$$

where $S_{\theta_{k_l}}(k_{l,t+1}, t+1)$ represents the LM function, responsible for capturing finer details during the denoising process. The noise is gradually reduced through multiple iterations, ultimately producing a result that is close to the original data. After applying the prediction and correction steps of the two-stage model, we also introduce a series of data processing operations to ensure the consistency and physical rationality of the reconstructed data. In particular, we use LR operators and DC optimization strategies to correct outliers that may arise from noise or prediction errors. The LR constraint unit constructs the Hankel matrix based on prior information generated in the k-space domain, and uses hard threshold singular value analysis of the Hankel matrix to recover the target contrast data from the under-sampled data. The LR operator optimization process is described as follows:

$$\min_{k_{l,t}} \| Ak_{l,t} - y \|_2^2 \ s.t. rank(L) = a, k_{l,t} = H^+(L), \quad (11)$$

where $H^+(\cdot)$ is the Hankel pseudo-inverse operator, which represents the data matrix after hard threshold singular value analysis. $L$ is a data matrix with low-rank property after conducting hard-threshold singular values operation. $a$ is the rank of the data matrix. Then, the sampled values are restored using DC. It can be expressed as:

$$k^* = \min_k \{\|Ak - y\|_2^2 + \lambda \|k - k_{l,t}\|_2^2\}, \quad (12)$$

where $k^*$ is the reconstruction result obtained at each iteration. According to the above formula, the corresponding DC solution can be solved by mathematical reasoning:

$$k(w) = \begin{cases} k^*(w) & if \ w \notin \Omega \\ \frac{A^T y + \lambda k^*(w)}{1+\lambda} & if \ w \in \Omega \end{cases}, \quad (13)$$

where $\Omega$ represents the index set of the acquired k-space samples, with $k(w)$ representing the entry at index $w$ in the k-space domain generated by the network. In a noise-free scenario (i.e., $\lambda \to \infty$), if coefficients at step $w$ have been sampled, the initial coefficients replace the predicted ones.

In summary, the combination of the inverse diffusion predictor and the Langevin dynamics corrector effectively reduces noise and generates high-quality MRI data throughout the reconstruction process. LR and DC steps serve to further improve the reliability and authenticity of the reconstructed image. This multi-level optimization strategy provides strong support for high-quality MRI reconstruction. Further details on the reconstruction algorithm can be found in Algorithm 1.

---
**Algorithm 1: GLDM**
---
**Training stage**
**1. Input:** k-space datasets: $k_g(t), k_l(t)$
**2. Training:** Eqs. (6), (8)
**3. Output:** $S_{\theta_{k_g}}(k_g(t), t), S_{\theta_{k_l}}(k_l(t), t)$
**Reconstruction stage**
**Setting:** $T, J, \varepsilon, \sigma$
1: $k^l \sim \mathbb{N}(0, \sigma_T^2 I)$
2: **for** $t = T - 1$ to 0 **do (Outer loop)**
3:   $k_{g,t} \leftarrow$ Predictor($k_{g,t+1}, \sigma_t, \sigma_{t+1}$)
4:   **for** $j = 1$ to $J$ **do (Inner loop)**
5:     $k_{g,t}^j \leftarrow$ Corrector($k_{g,t}^{j-1}, \sigma_t, \varepsilon_t$)
6:   **end**
7:   $k_{l,t} \leftarrow$ Predictor($k_{g,t}^J, \sigma_t, \sigma_{t+1}$)
8:   **for** $j = 1$ to $J$ **do (Inner loop)**
9:     $k_l^j \leftarrow$ Corrector($k_{l,t}^{j-1}, \sigma_t, \varepsilon_t$)
10:  **end**
11:  Update $k_l^j$ via Eqs. (11)-(12) to obtain $k^*$
12: **end**
13: **Return** $k \leftarrow k^*$
---

## IV. EXPERIMENTS

This section compares the performance of GLDM with distinctive types of methods under different acceleration factors and sampling modes. To ensure the comparability and fairness of the experiments, all methods were conducted on the same dataset. The source code of GLDM is accessible at: *https://github.com/yqx7150/GLDM.*

### A. Experiment Setup

***Data acquisition***: The 210 2D dynamic fully-sampled cardiac MR data used in the experiment are provided by the Shenzhen Institutes of Advanced Technology, Chinese Academy of Sciences. Specifically, all data are acquired on a 3.0T scanner (SIEMENS EQUIPTOM Trio) equipped with 20 coils using a balanced steady-state free precession (BSSFP) sequence for 30 volunteers. The BSSFP scan uses the following parameters: FOV 330 × 330 mm², acquisition matrix 256 × 256, slice thickness = 6 mm, TR/TE = 3.0 ms/1.5 ms. The time resolution is 40.0 ms. Each volunteer needs to hold his breath for 15-20 seconds on each slice. Each scan covers the entire cardiac dynamic process with one slice, with 25 time frames. The scanning parameters of all patients remain the same. It is worth noting that although fully-sampled k-space data can be obtained, they will not be used for model training, they are only used to simulate accelerated acquisition of under-sampled data. The specific time-interleaved accelerated acquisition scheme used has been described in detail in Section III-B. In the experiment, we first use the adaptive coil combination method to combine the original multi-coil data in the format of 192×192×16×20 ($x \times y \times k_t \times coil$) for 210 2D dynamic fully-sampled cardiac MR data to generate single-coil complex-valued data. Then, 200 single-coil complex-valued data are processed by time-interleaved accelerated acquisition, and finally 3200 under-sampled datasets in the format of 192×192 are obtained for model training. In addition, the remaining 10 single-coil complex-valued data are under-sampled using radial and cartesian masks, respectively, and used to reconstruct each frame.

***Implementation details***: For quantitative evaluation, we use Peak Signal-to-Noise Ratio (PSNR), Mean Square Error (MSE), and Structural Similarity Index (SSIM) to assess the quality of the reconstructed images. The proposed algorithm is trained in the k-space domain, the batch size is set to 4, and the Adam optimizer is used for training, where $\beta_1 = 0.9$ and $\beta_2 = 0.999$ are set to optimize the network. For the noise variance schedule, we set the parameters as $\sigma_{max} = 378$, $\sigma_{min} = 0.01$, and $r = 0.075$. By default, we use T = 300 iterations and J = 1 iteration for inference, unless otherwise specified. Furthermore, during the reconstruction phase, $\sigma_{max} = 4$, $\sigma_{min} = 0.01$ and $r = 0.075$ are set. For other parameters involved in the experiment, we set them according to the guidelines of the original paper [48]. Additionally, implemented on an Ubuntu 14.04 LTS (64-bit) operating system. The system is equipped with an Intel Core i7-4790K CPU and an NVIDIA GTX1080 GPU with 8 GB of memory. The implementation is carried out using the open-source framework of PyTorch, with CUDA and CUDNN support.

TABLE I
THE AVERAGE PSNR, SSIM, AND MSE ($*10^{-4}$) OF COMPARISON METHOD AND THE PROPOSED METHOD ON THE SINGLE-COIL CARDIAC CINE TEST DATASET WITH DIFFERENT ACCELERATIONS (8-FOLD AND 10-FOLD) AT RADIAL SAMPLING PATTERN.

| | | Method | 201 | 202 | 203 | 204 | 205 | 206 | 207 | 208 | 209 | 210 | Average |
|---|---|---|---|---|---|---|---|---|---|---|---|---|---|
| | PSNR | | 30.35 | 30.25 | 24.49 | 25.23 | 28.25 | 28.24 | 27.76 | 22.31 | 24.61 | 26.86 | 26.83 |
| | SSIM | BCS [49] | 0.7840 | 0.7754 | 0.7206 | 0.7299 | 0.6442 | 0.7074 | 0.6076 | 0.7026 | 0.7041 | 0.7410 | 0.7116 |
| | MSE | | 10.53 | 11.33 | 31.69 | 22.73 | 17.67 | 18.43 | 20.65 | 42.45 | 28.36 | 23.91 | 22.77 |
| | | | 30.99 | 29.93 | 26.26 | 27.34 | 28.41 | 27.72 | 28.02 | 24.89 | 26.61 | 27.73 | 27.79 |
| | | 3D-CSC [53] | 0.8200 | 0.7903 | 0.7597 | 0.7597 | 0.7461 | 0.7364 | 0.7109 | 0.7493 | 0.7503 | 0.7716 | 0.7594 |
| | | | 7.97 | 10.17 | 25.34 | 21.26 | 15.47 | 18.32 | 16.11 | 37.73 | 23.45 | 17.70 | 19.35 |
| | | | 31.18 | 30.19 | 25.56 | 26.99 | 28.06 | 27.25 | 27.63 | 23.98 | 26.08 | 26.36 | 27.32 |
| | | TGVNN [50] | 0.8548 | **0.8311** | 0.7688 | 0.7760 | **0.7734** | **0.7678** | **0.7607** | 0.7482 | **0.7774** | 0.7907 | 0.7848 |
| | | | 7.93 | 9.83 | 30.36 | 21.35 | 16.19 | 20.16 | 18.54 | 45.62 | 26.32 | 24.87 | 22.11 |
| **Radial** | | | 31.13 | 29.76 | 28.57 | 28.38 | 29.28 | 28.38 | 28.37 | 27.03 | 27.91 | 28.47 | 28.72 |
| **R = 8** | | DLTG [51] | 0.8233 | 0.7863 | 0.7690 | 0.7661 | 0.7461 | 0.7408 | 0.7162 | 0.7660 | 0.7537 | 0.7735 | 0.7641 |
| | | | 7.76 | 10.66 | 14.44 | 15.44 | 12.05 | 15.36 | 14.81 | 22.30 | 17.03 | 15.27 | 14.51 |
| | | | 30.56 | 27.32 | 27.83 | 28.95 | 28.24 | 28.69 | 25.30 | 27.31 | 27.81 | 29.61 | 28.16 |
| | | DD-UGM [52] | 0.8179 | 0.7518 | 0.7671 | 0.7603 | 0.7361 | 0.7188 | 0.7196 | 0.7342 | 0.7665 | 0.7708 | 0.7543 |
| | | | 8.84 | 19.81 | 18.38 | 13.08 | 16.39 | 13.68 | 34.35 | 20.06 | 17.56 | 11.28 | 17.34 |
| | | | **32.59** | **31.22** | **29.94** | **29.50** | **30.18** | **29.79** | **29.36** | **28.35** | **28.95** | **29.81** | **29.97** |
| | | GLDM | **0.8590** | 0.8279 | **0.7883** | **0.7874** | 0.7701 | 0.7621 | 0.7356 | **0.7771** | 0.7673 | **0.7940** | **0.7860** |
| | | | **5.52** | **7.57** | **10.81** | **12.04** | **9.69** | **11.00** | **11.79** | **17.11** | **13.99** | **11.24** | **11.07** |
| | PSNR | | 29.13 | 28.97 | 24.72 | 24.41 | 28.80 | 27.59 | 27.12 | 22.59 | 24.28 | 25.98 | 26.35 |
| | SSIM | BCS [49] | 0.7544 | 0.7614 | 0.7042 | 0.7136 | 0.6814 | 0.6254 | 0.5873 | 0.6894 | 0.6894 | 0.7278 | 0.6934 |
| | MSE | | 12.31 | 12.75 | 32.98 | 24.25 | 17.52 | 21.76 | 22.52 | 40.89 | 31.16 | 23.84 | 24.00 |
| | | | 31.29 | 30.14 | 25.35 | 26.98 | 28.26 | 27.49 | 27.75 | 24.04 | 25.75 | 26.85 | 27.39 |
| | | 3D-CSC [53] | 0.8350 | 0.8047 | 0.7574 | 0.7632 | 0.7526 | 0.7403 | 0.7180 | 0.7370 | 0.7500 | 0.7744 | 0.7632 |
| | | | 7.44 | 9.71 | 31.57 | 23.65 | 16.24 | 18.91 | 17.28 | 43.77 | 28.57 | 21.92 | 21.90 |
| | | | 30.36 | 29.52 | 24.36 | 25.97 | 27.32 | 26.81 | 26.96 | 23.04 | 25.05 | 25.84 | 26.52 |
| | | TGVNN [50] | 0.8400 | 0.8144 | 0.7432 | 0.7495 | 0.7536 | 0.7508 | **0.7420** | 0.7197 | 0.7545 | 0.7735 | 0.7641 |
| | | | 9.36 | 11.27 | 38.58 | 28.58 | 19.34 | 22.04 | 21.76 | 54.34 | 32.84 | 27.52 | 26.56 |
| **Radial** | | | 31.12 | 29.83 | 28.50 | 28.38 | 29.27 | 28.39 | 28.36 | 27.01 | 27.92 | 28.45 | 28.72 |
| **R = 10** | | DLTG [51] | 0.8232 | 0.7863 | 0.7690 | 0.7660 | 0.7460 | 0.7410 | 0.7162 | 0.7661 | 0.7539 | 0.7734 | 0.7594 |
| | | | 7.77 | 10.46 | 14.69 | 15.41 | 12.10 | 15.33 | 14.84 | 22.41 | 16.91 | 15.38 | 19.35 |
| | | | 31.44 | 30.01 | 27.42 | 27.71 | 28.61 | 27.79 | 28.17 | 25.73 | 26.75 | 27.83 | 28.14 |
| | | DD-UGM [52] | 0.8362 | 0.8021 | 0.7381 | 0.7449 | 0.7453 | 0.7196 | 0.7053 | 0.7107 | 0.7151 | 0.7538 | 0.7471 |
| | | | 7.19 | 10.01 | 18.51 | 18.10 | 14.20 | 17.35 | 15.47 | 30.79 | 22.63 | 17.49 | 17.17 |
| | | | **32.13** | **30.85** | **29.26** | **28.86** | **29.62** | **29.70** | **29.18** | **29.47** | **28.93** | **29.43** | **29.74** |
| | | GLDM | **0.8508** | **0.8164** | **0.7748** | **0.7751** | **0.7596** | **0.7530** | 0.7264 | **0.7709** | **0.7592** | **0.7823** | **0.7769** |
| | | | **6.15** | **8.24** | **12.82** | **14.33** | **11.08** | **10.89** | **12.22** | **11.68** | **13.92** | **12.05** | **11.34** |

## B. Reconstruction Comparison

To validate the advantages of GLDM, we present quantitative and qualitative comparisons with five representative dynamic MRI methods. It should be emphasized that these comparison algorithms either rely on fully-sampled data for model training or serve as traditional benchmarks to evaluate their performance without training. These methods include:

**BCS [49]**: A Bayesian CS-based reconstruction method as described in [49]. It applies low-order constraint techniques to improve reconstruction accuracy.

**TGVNN [50]**: A total generalized variation-based method, as described in [50]. It uses total variation regularization combined with neural networks for dynamic MRI reconstruction.

**DLTG [51]**: A DL-based tensor graph algorithm, described in [51], that leverages multi-scale features for dynamic MRI reconstruction.

**DD-UGM [52]**: A deep unfolding gradient descent method, described in [52], which incorporates unrolled gradient descent iterations with learned models for MRI reconstruction.

**3D-CSC [53]**: A 3D convolutional sparse coding algorithm, described in [53], which exploits sparse representations in three-dimensional dynamic MRI data.

For data processing, we mainly use the radial sampling scheme where each acquired row contains equal amounts of low- and high-frequency information, which helps reduce motion artifacts. Notably, all quantitative results are averaged over a 16-frame test dataset. Table I presents the quantitative evaluation of the reconstruction results of the above methods using acceleration factors $R = 8$ and $R = 10$. It can be seen that the proposed GLDM method generally outperforms other methods in the PSNR, SSIM, and MSE evaluations, which validates the effectiveness of our proposed method. Specifically, the reconstruction performance of GLDM improves PSNR by approximately 2 dB compared to the BCS and 3D-CSC methods, while the performance of the DLTG and DD-UGM methods is intermediate. On some datasets, TGVNN produces higher SSIM values than our algorithm, probably due to setting finer reconstruction parameters. Thus, GLDM method preserves high-quality image reconstruction and structural integrity without the necessity of fully-sampled data, exhibiting robust and adaptable performance across different acceleration and sampling patterns.

Figs. 4 and 5 illustrate the reconstruction results and the corresponding error plots of the distinct reconstruction techniques under 8-fold and 10-fold acceleration, respectively. To better understand the performance of each method in the time dimension, we also provide y-t images, i.e., the 100th slice extracted along the y-axis and the time dimension. It is clear from these plots that even at higher accelerations, our proposed method provides comparable or better reconstruction results than the other methods. In particular, our method significantly reduces the error in regions of interest such as the papillary muscle edges, whereas the other compared methods show larger residual errors in these regions. This suggests that GLDM is more effective in maintaining image quality, especially structural fidelity in critical regions of the image, under high acceleration conditions.

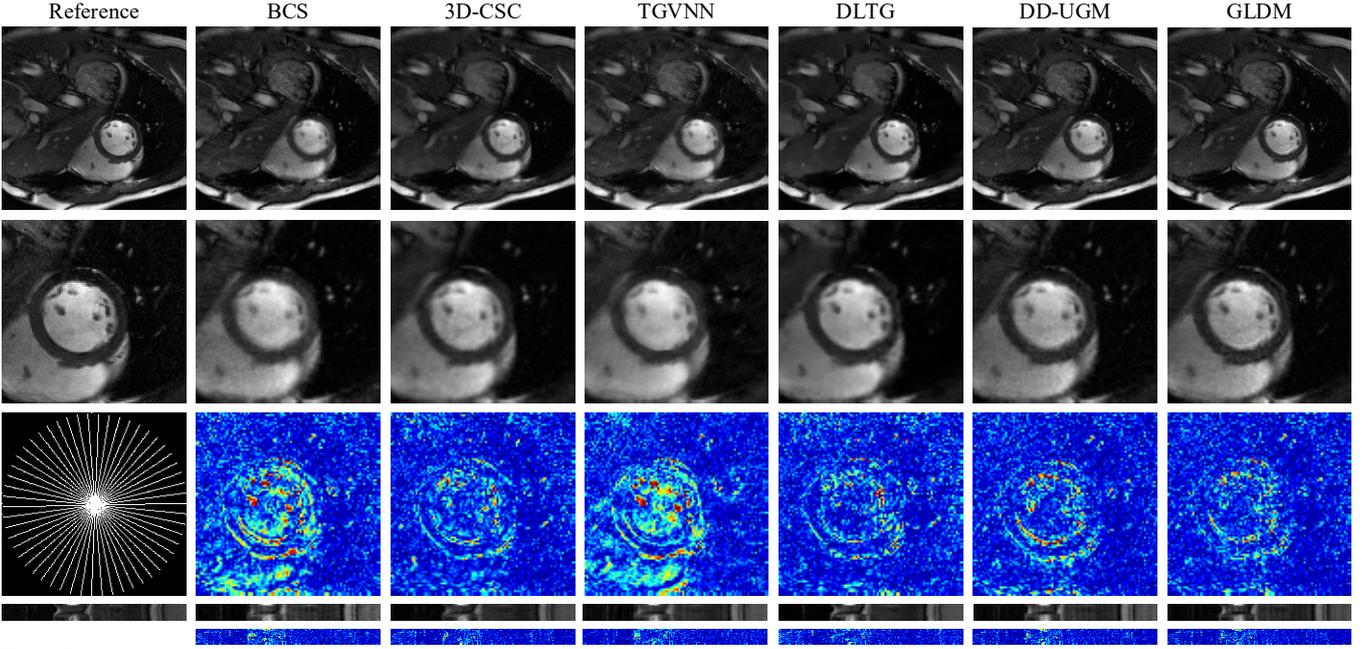

**Fig. 4.** Reconstruction results using BCS, 3D-CSC, TGVNN, DLTG, DD-UGM, and GLDM on the single-coil cardiac cine test 202 dataset with 8-fold acceleration and radial sampling. The second row highlights the region of interest, with the intensity of the residual maps magnified by a factor of five.

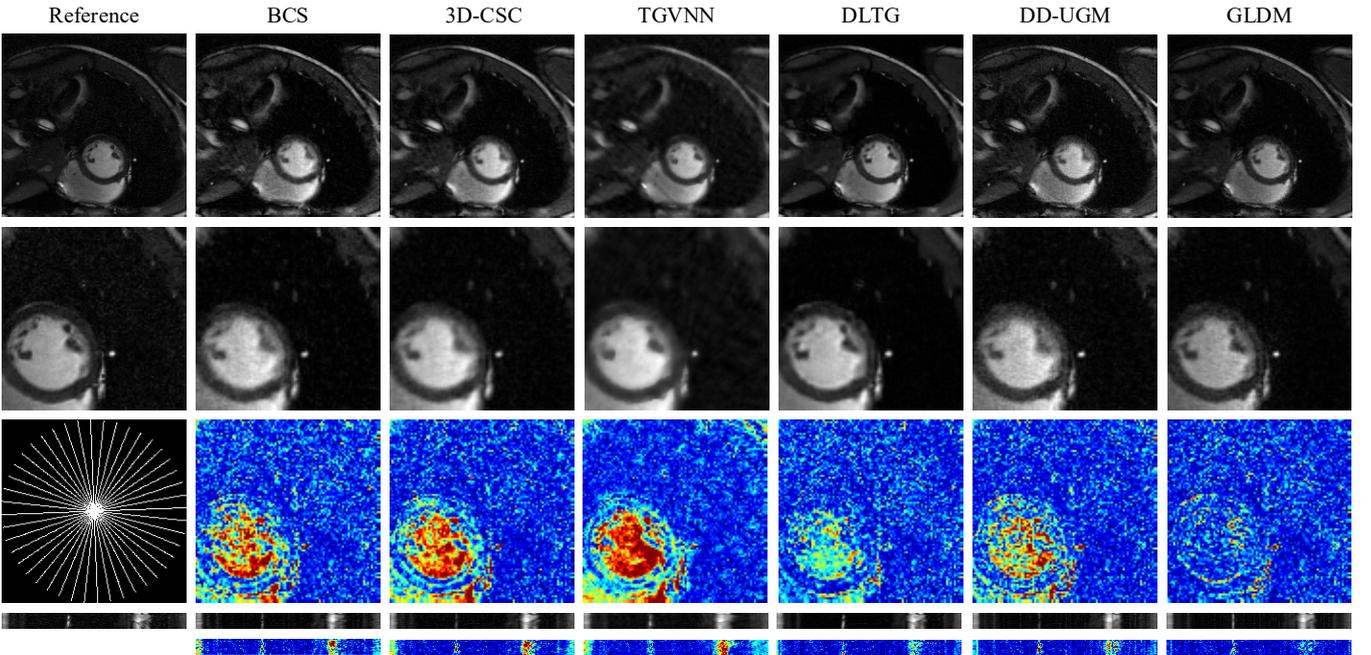

**Fig. 5.** Reconstruction results using BCS, 3D-CSC, TGVNN, DLTG, DD-UGM, and GLDM on the single-coil cardiac cine test 208 dataset with 10-fold acceleration and radial sampling. The second row highlights the region of interest, with the intensity of the residual maps magnified by a factor of five.

*C. Comparison of Sampling Schemes*

In this section, we compare the reconstruction results under the cartesian sampling scheme. Traditional MRI reconstruction methods require specific sampling distributions and strict sparsity conditions for the sampling mask, whereas DL-based methods impose no such stringent requirements. Although our proposed framework is based on a time-interleaved acquisition scheme, it is flexible and can accommodate any sampling scheme for both training and testing. Table II illustrates the reconstruction results with 6-fold and 8-fold acceleration under the cartesian sampling. Table II shows the quantitative performance comparison results of different reconstruction methods under the cartesian sampling scheme, which shows that GLDM has a significant advantage in reconstruction quality compared with k-t SLR [8] Specifically, GLDM shows better reconstruction results in both SSIM and PSNR, especially maintaining a low error level under high acceleration conditions, which proves its stability and robustness under different sampling acceleration conditions. Furthermore, the visualization of the cartesian mask is shown in Fig. 6. By comparing the reconstructed images and their corresponding error plots, it is evident that GLDM method achieves improved reconstruction results. These findings suggest that GLDM also exhibits strong generalization capabilities across different sampling schemes, achieving satisfactory reconstruction results.

TABLE II
THE AVERAGE PSNR, SSIM, AND MSE (*10⁻⁴) OF THE PROPOSED METHOD AT $R = 10$ AND $R = 8$ ON THE SINGLE-COIL CARDIAC CINE TEST DATASET UNDER CARTESIAN SAMPLING PATTERNS.

| Mask | Method | 201 | 202 | 203 | 204 | 205 | 206 | 207 | 208 | 209 | 210 | Average |
|---|---|---|---|---|---|---|---|---|---|---|---|---|
| Cartesian $R = 8$ | k-t SLR [8] | 31.14 | 29.51 | 25.85 | 27.84 | 28.74 | 28.05 | 28.03 | 25.62 | 26.85 | 27.40 | 27.90 |
| | | 0.8362 | 0.8055 | 0.7660 | 0.7718 | 0.7571 | 0.7577 | 0.7378 | 0.7611 | 0.7574 | 0.7808 | 0.7731 |
| | | 7.71 | 11.35 | 27.21 | 19.08 | 14.26 | 16.45 | 16.16 | 30.79 | 21.87 | 19.39 | 18.43 |
| | GLDM | **32.35** | **30.71** | **28.81** | **28.59** | **29.29** | **30.10** | **29.18** | **28.53** | **29.05** | **29.17** | **29.58** |
| | | **0.8693** | **0.8328** | **0.7808** | **0.7849** | **0.7742** | **0.7735** | **0.7488** | **0.7669** | **0.7655** | **0.7907** | **0.7887** |
| | | **5.83** | **8.53** | **13.94** | **14.69** | **11.929** | **10.19** | **12.28** | **15.19** | **12.77** | **12.63** | **11.80** |
| Cartesian $R = 6$ | k-t SLR [8] | 31.85 | 30.58 | 26.64 | 28.56 | 29.37 | 28.48 | 28.62 | 26.29 | 26.99 | 27.80 | 28.52 |
| | | 0.8541 | 0.8242 | 0.7850 | 0.7916 | 0.7781 | 0.7731 | 0.7541 | 0.7782 | 0.7726 | 0.7975 | 0.7909 |
| | | 6.54 | 8.88 | 22.60 | 15.85 | 12.43 | 15.48 | 14.39 | 27.25 | 20.77 | 17.42 | 16.16 |
| | GLDM | **33.08** | **30.98** | **30.07** | **29.43** | **30.17** | **30.46** | **29.63** | **29.15** | **29.67** | **30.26** | **30.29** |
| | | **0.8817** | **0.8469** | **0.8058** | **0.8056** | **0.7943** | **0.7903** | **0.7646** | **0.7960** | **0.7874** | **0.8126** | **0.8085** |
| | | **4.94** | **8.31** | **10.62** | **12.63** | **9.80** | **9.10** | **11.04** | **13.51** | **11.50** | **9.97** | **10.14** |

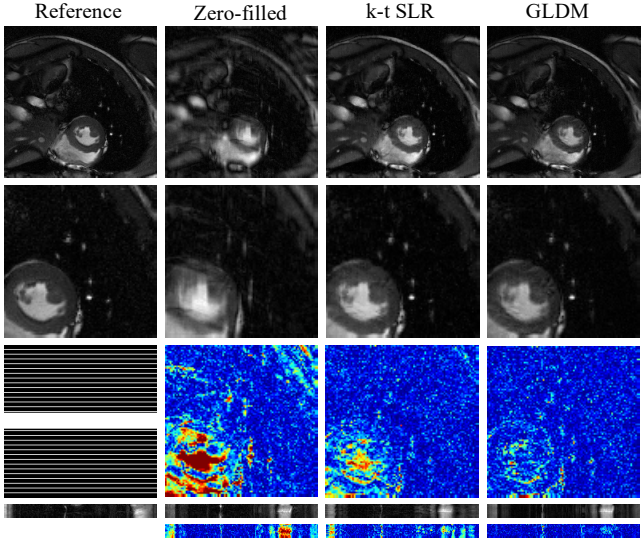

**Fig. 6.** Reconstruction results of k-t SLR and GLDM in the single-coil scenario under cartesian mask at 6-fold. The second row of images shows the region of interest, where the intensity of the residual map is magnified five times.

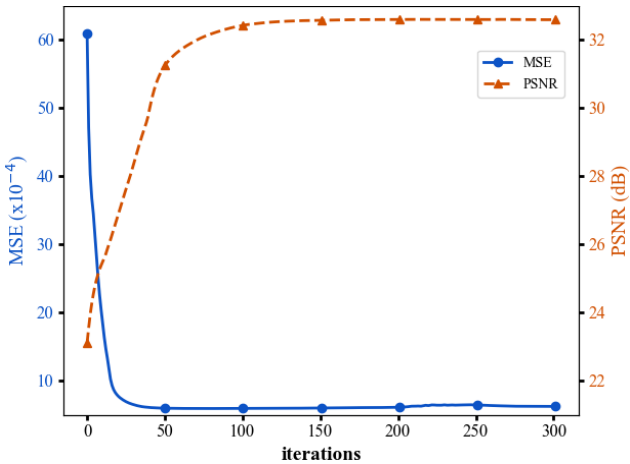

**Fig. 7.** Convergence curve of GLDM in terms of PSNR and MSE versus iterations.

### D. Iterative Convergence Analysis

Fig. 7 illustrates the convergence trend of GLDM algorithm across different iteration counts. In the first 100 iterations, PSNR rapidly increases from approximately 20dB to over 30dB, reflecting a significant improvement in image quality during the early optimization phase. Simultaneously, the MSE drops sharply from a high initial value to nearly below 10, indicating a substantial reduction in reconstruction error early on. As iterations continue, the rate of PSNR improvement slows and plateaus around 300 iterations, suggesting that the algorithm achieves convergence at this point, with PSNR stabilizing at a higher value. This implies that the optimal reconstruction quality has been reached. Similarly, the MSE curve shows a rapid decline in the early iterations, followed by a stable phase, demonstrating the algorithm's capacity to consistently reduce error. Overall, the graph highlights the effectiveness and stability of GLDM algorithm, with fast early improvements in both image quality and error reduction, followed by a convergence phase that confirms its robustness over multiple iterations.

## V. DISCUSSION

### A. Necessity of Local Prior Information

This section analyzes the role of LM in the reconstruction process. While GM effectively captures the overall structure and low-frequency components of the image, it often lacks the ability to adequately handle fine details and local features, resulting in poor reconstruction of complex regions. The LM addresses these deficiencies by learning the prior information required for more accurate reconstruction. As a result, the combination of both global and local models allows for a more comprehensive reconstruction process, leveraging the strengths of each to achieve higher-quality results.

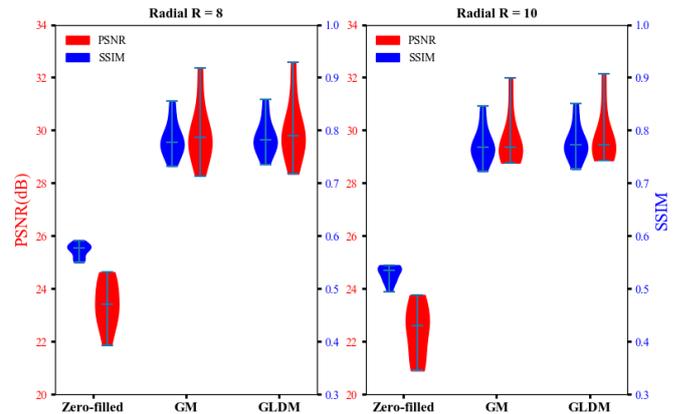

**Fig. 8.** Reconstruction performance of global model (GM) and GLDM under radial sampling with different acceleration factors.

As shown in Fig. 8, GLDM consistently outperforms GM in all 10 test samples although the improvement is not significant. Importantly, it is consistent across sampling modes with different acceleration factors, reflecting the benefit of incorporating local models during the reconstruction process. By integrating the LM, finer details that are typically smoothed over by GM are better preserved. This is particularly advantageous in regions with intricate textures or rapid temporal variations, where retaining such detail is crucial for accurate image reconstruction.

### B. Impact of Time-interleaved Acquisition Schemes

Although Section III-B has provided a detailed explanation of how the training dataset is obtained, it is important to note that the accelerated sampling based on the time-interleaved acquisition scheme is not restricted to this specific method. For example, Ke *et al.* [43] summed and averaged the adjacent frame data in the time direction. This method has some advantages, such as eliminating time redundancy and obtaining a higher signal-to-noise ratio. However, it also has some disadvantages, such as not fully exploiting the time correlation. In contrast, the proposed time-interleaved acquisition scheme solves these problems well and increases the utilization of the time correlation of adjacent frames. As shown in Fig. 9, the datasets constructed by the two different time-interleaved acquisition schemes are demonstrated, and it can be seen from the error plots that the dataset constructed by the proposed method is significantly better and closer to the fully-sampled data.

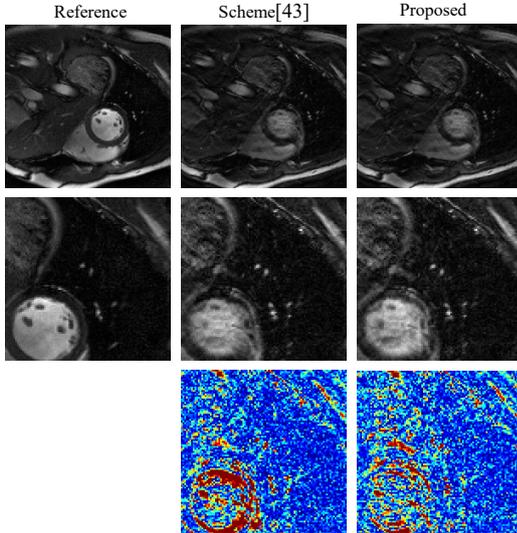

**Fig. 9.** Datasets constructed with two different time-interleaved acquisition schemes. The second row shows the region of interest and the error map is magnified five times
.

To further demonstrate the advantages of the proposed method, we utilized both schemes to construct datasets for training the model, ensuring that all other experimental settings and parameters are kept unchanged for the image reconstruction process. This allowed for a direct comparison between the two methods in terms of their effectiveness in reconstructing images. Table III shows the quantitative reconstruction results under the two methods with different samples. This suggests that the proposed method is more effective at capturing and reconstructing critical image details, leading to higher fidelity reconstructions.

TABLE III
PSNR, SSIM, AND MSE (*10$^{-4}$) OF DIFFERENT METHODS ON SINGLE-COIL CARDIAC CINE DATASET (MEAN ± STANDARD DEVIATION).

| Mask | Method | PSNR | SSIM | MSE |
|---|---|---|---|---|
| Cartesian $R = 8$ | Scheme [43] | 28.88±1.64 | 0.7653±0.0405 | 15.47±5.79 |
| | Proposed | **29.58±1.24** | **0.7887±0.0380** | **11.79±3.04** |
| Radial $R = 10$ | Scheme [43] | 28.16±1.64 | 0.7469±0.0395 | 15.42±5.97 |
| | Proposed | **29.74±1.00** | **0.7769±0.0346** | **11.33±2.63** |

### C. Impact of Frame Number on Prior Learning

This section studies the impact of different frame merging numbers on the performance of local models in prior learning. A brand new LM is trained by setting the number of merges to 8, which is combined with the already trained GM to construct a model called GLDM-8, while the LM used by the proposed algorithm is obtained by setting the number of merges to 4. As shown in Table IV, the quantitative reconstruction results of the two algorithms are demonstrated, and the comparative analysis shows that the proposed GLDM outperforms GLDM-8 in terms of quantitative reconstruction results under different modes of sampling methods. These findings further validate the effectiveness and optimality of the proposed model in enhancing prior learning.

TABLE IV
PSNR, SSIM, AND MSE (*10$^{-4}$) OF DIFFERENT METHODS ON SINGLE-COIL CARDIAC CINE DATASET (MEAN ± STANDARD DEVIATION).

| Mask | Method | PSNR | SSIM | MSE |
|---|---|---|---|---|
| Cartesian $R = 8$ | GLDM-8 | 28.93±1.66 | 0.7854±0.0394 | 14.42±5.72 |
| | GLDM | **29.58±1.24** | **0.7887±0.0380** | **11.79±3.04** |
| Radial $R = 10$ | GLDM-8 | 29.66±1.12 | 0.7760±0.0367 | 11.62±2.84 |
| | GLDM | **29.74±1.00** | **0.7769±0.0346** | **11.33±2.63** |

## VI. CONCLUSION

This paper proposed a novel generative model for dynamic MRI reconstruction based on fractional priors, leveraging the time-interleaved acquisition scheme. In particular, the proposed algorithm eliminated the need for fully-sampled data during training. Instead, it utilized the time-interleaved acquisition scheme to accelerate the acquisition of training data, efficiently exploiting the temporal redundancy inherent in dynamic MRI datasets. Concurrently, a two-stage learning framework, consisting of a global-to-local diffusion and an iterative optimization process using LR operators, enabled zero-shot reconstruction during the inference stage. Numerous experimental results demonstrated that the proposed algorithm achieved superior or comparable performance in both qualitative and quantitative evaluations. Further research focused on optimizing the time-interleaved acquisition scheme to improve its adaptability to a wider range of clinical applications. Additionally, the exploration of new prior models promised to further enhance the quality of reconstructed images.